\documentclass[12pt]{article}
\pdfoutput=1

\usepackage{amssymb}
\usepackage{amsmath}
\usepackage{amscd}
\usepackage{mathrsfs}
\usepackage{amssymb}
\usepackage{feynmf}
\usepackage[pdftex]{graphicx}
\usepackage{youngtab}

\def\dj{\hbox{d\kern-0.347em \vrule width 0.3em height 1.252ex depth
-1.21ex \kern 0.051em}}

\numberwithin{equation}{section}

\begin{document}

\setlength{\oddsidemargin}{0cm}
\setlength{\baselineskip}{7mm}


\thispagestyle{empty}
\setcounter{page}{0}

\begin{flushright}

\end{flushright}

\vspace*{1cm}

\begin{center}
{\bf \Large Planar Zeros in Gauge Theories and Gravity}

\vspace*{1cm}

Diego Medrano Jim\'enez$^{a,}$\footnote{\tt  d.medrano@csic.es}, 
Agust\'{\i}n Sabio Vera$^{a,}$\footnote{\tt 
a.sabio.vera@gmail.com}
and Miguel \'A. V\'azquez-Mozo$^{b,}$\footnote{\tt 
Miguel.Vazquez-Mozo@cern.ch}

\end{center}

\vspace*{0.0cm}

\begin{center}

$^{a}${\sl Instituto de F\'{\i}sica Te\'orica UAM/CSIC \&
Universidad Aut\'onoma de Madrid\\
C/ Nicol\'as Cabrera 15, E-28049 Madrid, Spain
}

$^{b}${\sl Instituto Universitario de F\'{\i}sica Fundamental y Matem\'aticas (IUFFyM) \\
 Universidad de Salamanca \\ 
 Plaza de la Merced s/n,
 E-37008 Salamanca, Spain
  }
\end{center}

\vspace*{2cm}

\centerline{\bf \large Abstract}

\noindent
Planar zeros are studied in the context of the five-point scattering amplitude for gauge bosons and gravitons. 
In the case of gauge theories, it is found that planar zeros are determined by an algebraic curve in the projective plane spanned
by the three stereographic coordinates labelling the direction of the outgoing momenta. This curve depends on the values of six
independent color structures. Considering the gauge group SU($N$) with $N=2,3,5$ and fixed color indices,
the class of curves obtained gets broader by increasing the rank of the group. For the five-graviton scattering, on the other hand,
we show that the amplitude vanishes whenever the process is planar, without imposing further kinematic conditions.
A rationale for this result is provided using color-kinematics duality.

\newpage  

\setcounter{footnote}{0}

\section{Introduction}

In the last decade, many studies have permitted a deeper understanding of the relationship between gravity and gauge theories from the point of view of scattering amplitudes (see \cite{review} for a comprehensive review). One of the most interesting results is color-kinematics
duality \cite{color-kinematics}, which allows the construction of gravity amplitudes by replacing color factors by a second copy of the 
kinematic numerators. This double copy structure has a historic antecedent in the Kawai, Lewellen, and Tye (KLT) relations \cite{KLT},
showing how, at tree level, closed string amplitudes admit a decomposition in terms of products of open string amplitudes. Similar
structures have been found in various other setups \cite{stieberger_taylor,ours}. It seems clear that, at
the level of scattering amplitudes, there is a sense in which gravity can be considered the ``square'' of a gauge theory.  

Given this double copy structure, a natural question to ask is how certain properties of gauge theory amplitudes translate 
into the gravitational side. One of these is the existence of radiation zeros \cite{rad_zeros_rev}. This is a peculiar 
feature of certain scattering processes where one or more massless gauge bosons are radiated, consisting in the vanishing of the amplitude
for certain phase space configurations. 
The phenomenon was first identified in processes involving gauge bosons trilinear couplings, in particular 
$u\bar{d}\rightarrow W^{+}\gamma$ \cite{rad-zeros}. It has been experimentally observed both at the Tevatron \cite{tevatron_zeros}
and LHC \cite{LHC_zeros}. Their existence has been also studied in graviton photoproduction \cite{passarino}. 

These so-called Type-I zeros appear for momentum configurations satisfying the constraint $Q_{i}=\kappa \,  p_{i}\cdot k$, where $k$ is the
momentum of the gauge boson, $Q_{i}$ and $p_{i}$ are the charge and momenta of the other particles, and $\kappa$ is a numerical constant.
In Ref. \cite{heyssler_stirling} it was realized that zeros in the amplitude may also occur when the spatial momenta of the
particles involved in the process lie on the same plane. These Type-II or planar zeros have been  identified
in the processes $e^{+}e^{-}\rightarrow W^{+}W^{-}\gamma$ \cite{soft1} 
and $e^{+}e^{-}\rightarrow \tau\bar{\tau}\gamma$ \cite{soft2}, in both cases in the soft photon limit. 

So far, the only study of planar zeros beyond the soft limit has been carried out in the interesting work  \cite{harland-lang}, where the five parton amplitude
in QCD was analyzed. Using the maximally helicity violating (MHV) formalism, planar zeros were found both for the $gg\rightarrow ggg$ and
$q\bar{q}\rightarrow ggg$ processes. 
In the case of the five-gluon amplitude for general color factors the planar zero condition depends on the color quantum numbers of incoming
and outgoing gluons. 

The present article has a double aim. One is to study the conditions for the emergence of planar zeros in the five-gluon amplitude. 
We show that planar zeros are a ``projective" property of the amplitude, in the sense that they are preserved by a simultaneous rescaling of the stereographic coordinates labeling the flight directions of the three outgoing gluons.
In terms of stereographic coordinates, we find that the existence of planar zeros is determined by a cubic algebraic curve whose integer coefficients are
given in terms of the color factors. In the case of SU($N$) gauge groups, 
we find that the casuistic of curves obtained for different color configurations gets broader as the rank $N$ increases, 
starting with the case of SU(2) where no physical zeros are found for external particles with well-defined color quantum numbers. 
Our second target consists in exploiting color-kinematics duality to study planar zeros in the gravitational case, where
we find that the five-graviton amplitude vanishes whenever the process is planar.  
This can be understood applying the BCJ prescription to the equation determining the zeros in the gauge case. By replacing
color factors with kinematic numerators satisfying color-kinematics duality, 
the condition for the planar zero is seen to be identically satisfied without further kinematic
constraints.  

The plan of the paper is as follows. In Section \ref{sec:5-gluon} we review the calculation of the five gluon amplitude
using the MHV formalism. Section \ref{sec:gluon_planar} is devoted to the conditions for planar zeros in the gauge case, 
while in Section \ref{sec:permutations} we study the transformation of the loci of planar zeros under permutations of the color labels of the external gluons. In Section \ref{sec:graviton} the graviton amplitude is obtained using color-kinematics duality and the condition
for the existence of amplitude zeros is obtained. Finally, in Section \ref{sec:conclusions} we summarize our conclusions.

\section{The five-gluon amplitude}
\label{sec:5-gluon}

In this section we revisit the construction of the five-gluon amplitude 
\begin{align}
g(p_{1},a_{1})+g(p_{2},a_{2})\longrightarrow g(p_{3},a_{3})+g(p_{4},a_{4})+g(p_{5},a_{5}),
\end{align}
where we take all momenta incoming. The tree level amplitude
is computed in terms of 15 nonequivalent trivalent diagrams, leading to the expression
\begin{align}
\mathcal{A}_{5}&=g^{3}\left({c_{1}n_{1}\over s_{12}s_{45}}+{c_{2}n_{2}\over s_{23}s_{15}}+{c_{3}n_{3}\over s_{34}s_{12}}
+{c_{4}n_{4}\over s_{45}s_{23}}+{c_{5}n_{5}\over s_{15}s_{34}}+{c_{6}n_{6}\over s_{14}s_{25}}
+{c_{7}n_{7}\over s_{13}s_{25}}+{c_{8}n_{8}\over s_{24}s_{13}} \right. \nonumber \\[0.2cm]
&+ \left.{c_{9}n_{9}\over s_{35}s_{24}}+{c_{10}n_{10}\over s_{14}s_{35}}+{c_{11}n_{11}\over s_{15}s_{24}}
+{c_{12}n_{12}\over s_{12}s_{35}}+{c_{13}n_{13}\over s_{23}s_{14}}+{c_{14}n_{14}\over s_{25}s_{34}}
+{c_{15}n_{15}\over s_{13}s_{45}}\right),
\label{eq:amplitude_gluon_general}
\end{align}
where we have introduced the kinematic invariants
\begin{align}
s_{ij}=(p_{i}+p_{j})^{2}=2 \, p_{i}\cdot p_{j}, \hspace*{1cm} i<j.
\end{align}
The color factors in Eq. \eqref{eq:amplitude_gluon_general} are given by\footnote{Our conventions for the 
color factors differ from those in Refs. \cite{review,color-kinematics}.}  
\begin{align}
c_{1} &= f^{a_{1}a_{2}b} f^{ba_{3}c} f^{ca_{4}a_{5}}, \hspace*{1cm} c_{2}=f^{a_{1}a_{5}b} f^{ba_{4}c} f^{ca_{3}a_{2}},
\nonumber \\[0.2cm]
c_{3} &= f^{a_{3}a_{4}b} f^{ba_{5}c} f^{ca_{1}a_{2}}, \hspace*{1cm} c_{4}=f^{a_{4}a_{5}b} f^{ba_{1}c} f^{ca_{2}a_{3}},
\nonumber \\[0.2cm]
c_{5} &= f^{a_{5}a_{1}b} f^{ba_{2}c} f^{ca_{3}a_{4}}, \hspace*{1cm} c_{6}=f^{a_{1}a_{4}b} f^{ba_{3}c} f^{ca_{5}a_{2}},
\nonumber \\[0.2cm]
c_{7} &= f^{a_{1}a_{3}b} f^{ba_{4}c} f^{ca_{5}a_{2}}, \hspace*{1cm} c_{8}=f^{a_{1}a_{3}b} f^{ba_{5}c} f^{ca_{4}a_{2}},
\label{eq:color_factors} \\[0.2cm]
c_{9} &= f^{a_{3}a_{5}b} f^{ba_{1}c} f^{ca_{2}a_{4}}, \hspace*{1cm} c_{10}=f^{a_{4}a_{1}b} f^{ba_{2}c} f^{ca_{3}a_{5}},
\nonumber \\[0.2cm]
c_{11} &= f^{a_{1}a_{5}b} f^{ba_{3}c} f^{ca_{4}a_{2}}, \hspace*{1cm} c_{12}=f^{a_3a_5b} f^{ba_4c} f^{ca_1a_2},
\nonumber \\[0.2cm]
c_{13} &= f^{a_{1}a_{4}b} f^{ba_{5}c} f^{ca_{3}a_{2}}, \hspace*{1cm} c_{14}=f^{a_{5}a_{2}b} f^{ba_1c} f^{ca_{3}a_{4}},
\nonumber \\[0.2cm]
c_{15} &= f^{a_{1}a_{3}b} f^{ba_{2}c} f^{ca_{4}a_{5}},
\nonumber
\end{align}
and satisfy nine independent Jacobi identities
\begin{align}
c_{3}-c_{5}+c_{14}&=0, \hspace*{1.15cm} c_{3}-c_{1}-c_{12}=0, \nonumber \\[0.2cm]
c_{4}-c_{1}+c_{15}&=0, \hspace*{1.15cm} c_{4}+c_{2}-c_{13}=0, \nonumber \\[0.2cm]
c_{5}+c_{2}-c_{11}&=0, \hspace*{1cm} c_{13}-c_{6}+c_{10}=0,
\label{eq:jacobi_id}\\[0.2cm]
c_{14}-c_{7}+c_{6}&=0, \hspace*{1.15cm} c_{7}-c_{8}+c_{15}=0, \nonumber \\[0.2cm]
c_{8}-c_{9}-c_{11}&=0, \hspace*{0.85cm} (c_9-c_{10}+c_{12}=0).
\nonumber
\end{align}

\begin{figure}[t]
\centerline{\includegraphics[scale=0.37]{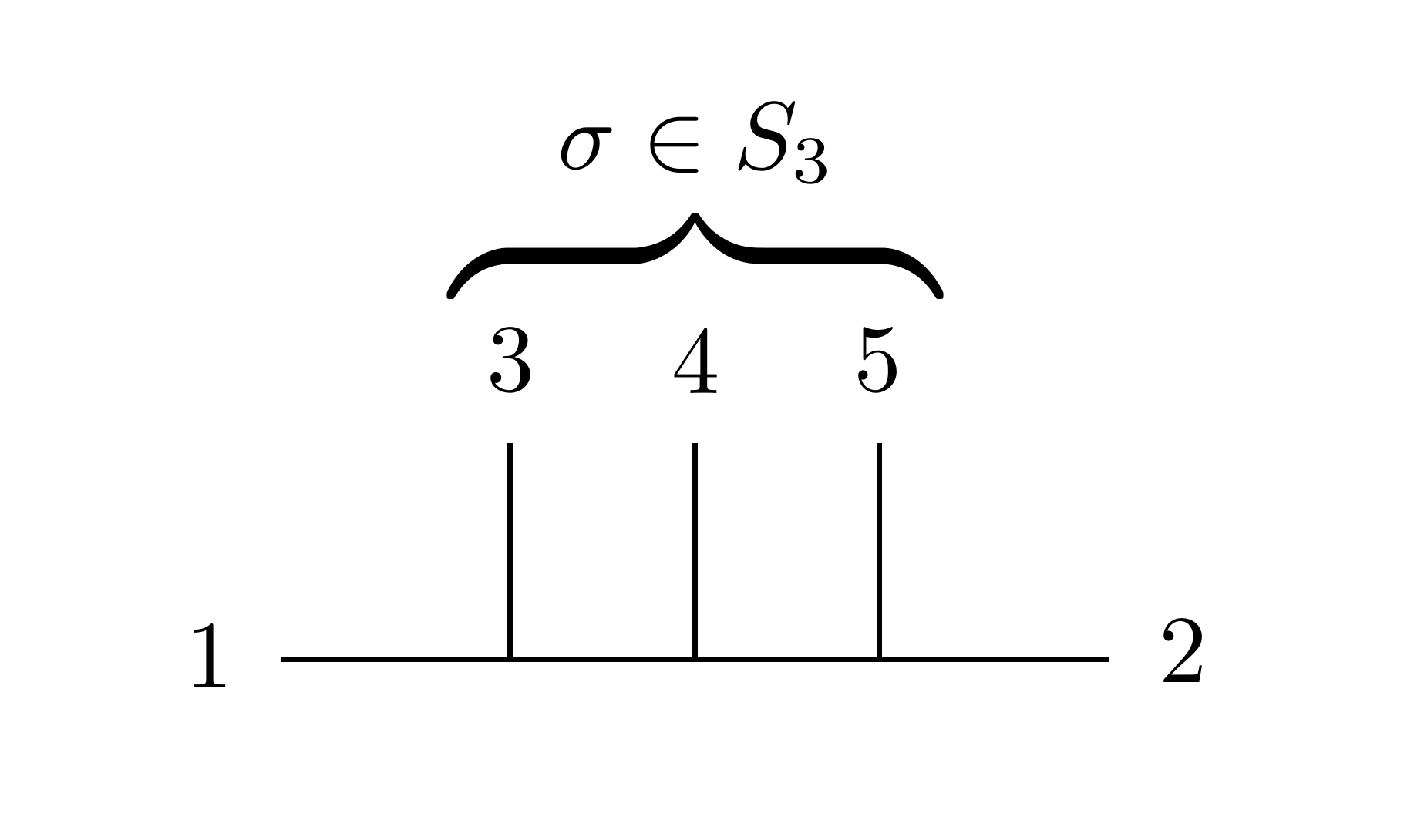}}
\caption[]{Graphic representation of the choice of subamplitudes basis in the implementation of the Kleiss-Kuijf relations.}
\label{fig:KK-rel}
\end{figure}

On general grounds, the amplitude can be written in terms of color-ordered amplitudes as
\begin{align}
\mathcal{A}_{5}=g^{3}\sum_{\sigma\in S_{4}}c[1,\sigma(2,3,4,5)]A_{5}[1,\sigma(2,3,4,5)],
\end{align}
where the sum is over noncyclic permutations of the external legs. However, the set of color-ordered amplitudes is over complete,
a fact expressed by the Kleiss-Kuijf relations 
\cite{kleiss_kuijf}. In the case of the five-point amplitude, there are $5\times 4$ different ways of choosing a basis in the space of 
independent color structures $\mathsf{TCS}_{5}$ \cite{kol_shir}. We select one of these basis by fixing the
incoming gluons (see Fig. \ref{fig:KK-rel}), 
so the five-gluon amplitude in Eq. \eqref{eq:amplitude_gluon_general} can be re-expressed in terms
of $3!$ color ordered amplitudes according to
\begin{align}
\mathcal{A}_{5}=g^{3}\sum_{\sigma\in S_{3}}c[1,2,\sigma(3,4,5)]A_{5}[1,2,\sigma(3,4,5)],
\label{eq:kleiss-kuijf_exp}
\end{align} 
where the subamplitudes are explictly given in terms of the numerators $n_{i}$ by
\begin{align}
A_{5}[1,2,3,4,5]&={n_{1}\over s_{12}s_{45}}-{n_{2}\over s_{23}s_{15}}+{n_{3}\over s_{34}s_{12}}+{n_{4}\over s_{45}s_{23}}
+{n_{5}\over s_{15}s_{34}}, \nonumber \\[0.2cm]
A_{5}[1,2,3,5,4]&=-{n_{1}\over s_{12}s_{45}}-{n_{13}\over s_{23}s_{14}}+{n_{12}\over s_{35}s_{12}}-{n_{4}\over s_{45}s_{23}}
+{n_{10}\over s_{14}s_{35}}, \nonumber \\[0.2cm]
A_{5}[1,2,4,3,5]&=-{n_{12}\over s_{12}s_{35}}-{n_{11}\over s_{24}s_{15}}-{n_{3}\over s_{34}s_{12}}+{n_{9}\over s_{35}s_{24}}
-{n_{5}\over s_{15}s_{34}}, \\[0.2cm]
A_{5}[1,2,4,5,3]&={n_{12}\over s_{12}s_{35}}-{n_{8}\over s_{24}s_{13}}-{n_{1}\over s_{45}s_{12}}-{n_{9}\over s_{35}s_{24}}
-{n_{15}\over s_{13}s_{45}}, \nonumber \\[0.2cm]
A_{5}[1,2,5,3,4]&=-{n_{3}\over s_{12}s_{34}}-{n_{6}\over s_{25}s_{14}}-{n_{12}\over s_{35}s_{12}}+{n_{14}\over s_{34}s_{25}}
-{n_{10}\over s_{14}s_{35}}, \nonumber \\[0.2cm]
A_{5}[1,2,5,4,3]&={n_{3}\over s_{12}s_{34}}-{n_{7}\over s_{25}s_{13}}+{n_{1}\over s_{12}s_{45}}-{n_{14}\over s_{34}s_{25}}
+{n_{15}\over s_{13}s_{45}}. \nonumber
\end{align}
Going back to the expression for the color factors in Eq. \eqref{eq:color_factors}, these partial amplitudes are respectively associated with
the six color factors $c_{7}$, $c_{8}$, $c_{6}$, $c_{13}$, $c_{11}$, and $c_{2}$. 

At this point we can exploit the generalized gauge freedom in the definition of the numerators to implement color-kinematics duality, so
the numerators $n_{i}$ mimic the Jacobi identities \eqref{eq:jacobi_id}. Solving the corresponding equations we can eliminate
$n_{7}$ to $n_{15}$ finding the following solution for the numerators in terms of the basis of color-ordered amplitudes
\begin{align}
n_{1}&=-n_{12}=n_{15}=s_{12}s_{45}A_{5}[1,2,3,4,5], \nonumber \\[0.2cm]
n_{2}&=n_{3}=n_{4}=n_{5}=n_{11}=n_{13}=n_{14}=0, \nonumber \\[0.2cm]
n_{6}&=n_{7}=n_{10}=s_{14}s_{35}A_{5}[1,2,3,5,4]+s_{14}(s_{35}+s_{45})A_{5}[1,2,3,4,5], \\[0.2cm]
n_{8}&=n_{9}=s_{14}s_{35}A_{5}[1,2,3,5,4]+(s_{14}s_{35}+s_{14}s_{45}+s_{12}s_{45})A_{5}[1,2,3,4,5].
\nonumber
\end{align}
Color-kinematics duality is independent of the polarization of the gluons. Here we are going to use the MHV formalism and assign negative 
helicity to the incoming gluons. Using the Parke-Taylor formula \cite{parke_taylor} we have
\begin{align}
A_{5}[1^{-},2^{-},\sigma(3^{+},4^{+},5^{+})]=i{\langle 12\rangle^{4}\over \langle 12\rangle\langle 2\sigma(3)\rangle 
\langle \sigma(3)\sigma(4)\rangle 
\langle \sigma(4)\sigma(5)\rangle\langle \sigma(5)1\rangle},
\end{align}
for any permutation $\sigma\in S_{3}$ of the last three indices. Expressing in addition the kinematic invariants in terms of spinors,
$s_{ij}=\langle ij\rangle[ji]$, we arrive at the following expressions for the numerators
\begin{align}
n_{1}&=-n_{12}=n_{15}=i{\langle 12\rangle^{4}[21][54]\over \langle 23\rangle\langle 34\rangle \langle 51\rangle}, \nonumber \\[0.2cm]
n_{6}&=n_{7}=n_{10}=i{\langle 12\rangle^{4}[14][52]\over \langle 23\rangle\langle 34\rangle \langle 51\rangle}, 
\nonumber \\[0.2cm]
n_{8}&=n_{9}=i{\langle 12\rangle^{4}[24][51]\over \langle 23\rangle\langle 34\rangle \langle 51\rangle}, \hspace*{1cm} 
\label{eq:numerators} \\[0.2cm]
n_{2}&=n_{3}=n_{4}=n_{5}=n_{11}=n_{13}=n_{14}=0.
\nonumber
\end{align}
With this result, the five-gluon amplitude can be written as
\begin{align}
\mathcal{A}_{5}&=-ig^{3}\langle 12\rangle^{3}\left({c_{2}\over \langle 23\rangle\langle 34\rangle\langle 45\rangle\langle 51\rangle}
+{c_{6}\over \langle 25\rangle\langle 53\rangle\langle 34\rangle\langle 41\rangle}
+{c_{7}\over \langle 25\rangle\langle 54\rangle\langle 43\rangle\langle 31\rangle}\right. \nonumber \\[0.2cm]
&\left.+{c_{8}\over \langle 24\rangle\langle 45\rangle\langle 53\rangle\langle 31\rangle} 
+{c_{11}\over \langle 24\rangle\langle 43\rangle\langle 35\rangle\langle 51\rangle} 
+{c_{13}\over \langle 23\rangle\langle 35\rangle\langle 54\rangle\langle 41\rangle}\right).
\label{eq:5g_amplitude_spinors}
\end{align}
Alternatively, this expression can be obtained from Eq. \eqref{eq:kleiss-kuijf_exp} by a direct application of the Parke-Taylor formula.

The spinor products appearing in the five-gluon amplitude can now be recast in terms of momenta. 
Working in the center-of-mass frame, the incoming momenta take the form
\begin{align}
p_{1}={\sqrt{s}\over 2}(1,0,0,1), \hspace*{1cm} p_{2}={\sqrt{s}\over 2}(1,0,0,-1).
\end{align}
On the other hand, for the three outgoing gluons their spatial momenta are parametrized using 
stereographic coordinates $\zeta_{a}\in \mathbb{C}$ (with $a=3,4,5$) according to
\begin{align}
p_{a}=-\omega_{a}\left(1,{\zeta_{a}+\overline{\zeta}_{a}\over 1+\zeta_{a}\overline{\zeta}_{a}},i{\overline{\zeta}_{a}-\zeta_{a}\over
1+\zeta_{a}\overline{\zeta}_{a}},{\zeta_{a}\overline{\zeta}_{a}-1\over 1+\zeta_{a}\overline{\zeta}_{a}}\right),
\end{align}
where the global minus sign reflects that all momenta are taken entering the diagram. The stereographic coordinates $\zeta_{a}$ 
are related to the rapidity $y_{a}$ and the azimuthal angle $\phi_{a}$ by
\begin{align}
\zeta_{a}=e^{y_{a}+i\phi_{a}}.
\end{align}

\section{Gauge planar zeros}
\label{sec:gluon_planar}

We focus now on planar five-gluon scattering with general color quantum numbers. Since the incoming particles travel along
the $z$ axis, without loss of generality we can take all momenta lying on the $xz$-plane. This means that $p_{a}^{y}=0$ and therefore 
$\zeta_{a}$ has to be real and the outgoing momenta read
\begin{align}
p_{a}=-{\omega_{a}\over 1+\zeta_{a}^{2}}(1+\zeta_{a}^{2},2\zeta_{a},0,\zeta_{a}^{2}-1).
\end{align}
Alternatively, the planarity condition implies that all emitted particles have azimuthal angles with either $\phi_{a}=0$ or $\phi_{a}=\pi$.

Implementing momentum conservation $p_{1}+\ldots+p_{5}=0$ gives three independent equations that determine the gluon energies $\omega_{a}$ in terms
of the center-of-mass energy $\sqrt{s}$ and the flight directions of the outgoing gluons labelled by $\zeta_{a}$,
\begin{align}
\omega_{3}&={\sqrt{s}\over 2}{(1+\zeta_{3}^{2})(1+\zeta_{4}\zeta_{5})\over (\zeta_{3}-\zeta_{4})(\zeta_{3}-\zeta_{5})}, 
\nonumber \\[0.2cm]
\omega_{4}&={\sqrt{s}\over 2}{(1+\zeta_{4}^{2})(1+\zeta_{3}\zeta_{5})\over (\zeta_{4}-\zeta_{3})(\zeta_{4}-\zeta_{5})},
\label{eq:energies_dehomog}\\[0.2cm]
\omega_{5}&={\sqrt{s}\over 2}{(1+\zeta_{5}^{2})(1+\zeta_{3}\zeta_{4})\over (\zeta_{5}-\zeta_{3})(\zeta_{5}-\zeta_{4})}.
\nonumber
\end{align}
Furthermore, the finite positive energy condition $0\leq \omega_{a}<\infty$ imposes constraints on the possible values of $\zeta_{a}$. In particular, let us remark that finite energy implies that $\zeta_{a}\neq \zeta_{b}$ for $3\leq a<b\leq 5$. 

Using this parametrization, the amplitude \eqref{eq:5g_amplitude_spinors} takes the form
\begin{align}
\mathcal{A}_{5}&={2ig^{3}\over\sqrt{s}}{(\zeta_{3}-\zeta_{4})(\zeta_{3}-\zeta_{5})(\zeta_{4}-\zeta_{5})\over
(1+\zeta_{3}\zeta_{4})(1+\zeta_{3}\zeta_{5})(1+\zeta_{4}\zeta_{5})}\left[-c_{2}{\zeta_{5}-\zeta_{3}\over \zeta_{3}}
-c_{6}{\zeta_{4}-\zeta_{5}\over \zeta_{5}}\right. \nonumber \\[0.2cm]
&\left.+c_{7}{\zeta_{3}-\zeta_{5}\over \zeta_{5}}-c_{8}{\zeta_{3}-\zeta_{4}\over \zeta_{4}}
+c_{11}{\zeta_{5}-\zeta_{4}\over\zeta_{4}}+c_{13}{\zeta_{4}-\zeta_{3}\over \zeta_{3}}
\right].
\label{eq:amplitude_zetas}
\end{align}
In order to find the zeros of the amplitude, we notice that the finite energy condition implies that the prefactor can never vanish.
As a consequence, we find the following equation depending on the color factors
\begin{align}
c_{2}{\zeta_{5}-\zeta_{3}\over \zeta_{3}}
+c_{6}{\zeta_{4}-\zeta_{5}\over \zeta_{5}}-c_{7}{\zeta_{3}-\zeta_{5}\over \zeta_{5}}
\hspace*{3cm}\nonumber \\[0.2cm]
+c_{8}{\zeta_{3}-\zeta_{4}\over \zeta_{4}}
-c_{11}{\zeta_{5}-\zeta_{4}\over\zeta_{4}}-c_{13}{\zeta_{4}-\zeta_{3}\over \zeta_{3}}=0.
\label{eq:planar_cond1}
\end{align}

The planar zero condition just derived is a homogeneous equation of vanishing degree. Since the amplitude \eqref{eq:amplitude_zetas}
diverges whenever any of the $\zeta_{a}$ vanishes, we can multiply the previous equation by $\zeta_{3}\zeta_{4}\zeta_{5}$
without generating spurious solutions in the physical region. Taking projective coordinates 
\begin{align}
(\zeta_{3},\zeta_{4},\zeta_{5})=\lambda(1,U,V), \hspace*{1cm} \lambda,U,V\neq 0 
\label{eq:projective_coord}
\end{align}
the planar zeros of the five-gluon amplitude are determined by the loci defined by 
the following equation
\begin{align}
c_{7}U-c_{8}V-c_{6}U^{2}+c_{11}V^{2}+(c_{2}+c_{6}-c_{7}+c_{8}-c_{11}-c_{13})UV+c_{13}U^{2}V-c_{2}UV^{2}=0.
\label{eq:cubic_eq}
\end{align}
Moreover, this equation is homogeneous in the color factors and therefore independent of the normalization of the
gauge group generators. Since there exists a normalization of the generators that makes all structure constants 
integer numbers \cite{sattinger-weaver}, the planar zeros are determined by a cubic curve with integer coefficients. 

\begin{figure}[t]
\centerline{\includegraphics[scale=0.5]{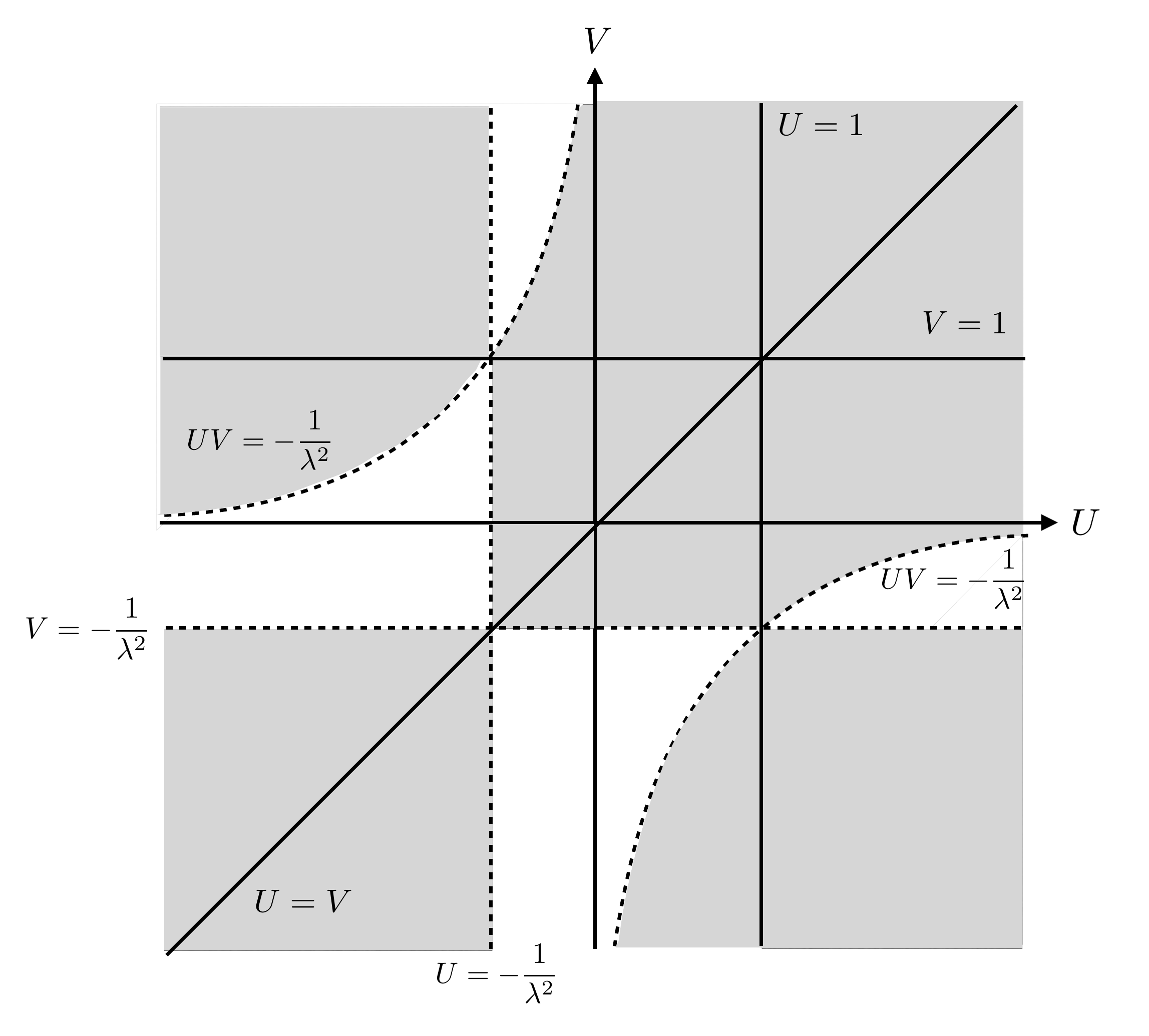}}
\caption[]{Physical regions in the $UV$ plane for a given value of $\lambda$. The shadowed regions are unphysical points where
the energy is negative for at least one of the outgoing gluons. Dashed lines correspond to the soft limits in 
which one or several energies tend to zero. Solid lines, including the axes, represent also unphysical points.}
\label{fig:physical_regions}
\end{figure}

In terms of the projective coordinates \eqref{eq:projective_coord}, the energies of the outgoing particles take the form
\begin{align}
\omega_{3}&={\sqrt{s}\over 2}{(1+\lambda^{2})(1+\lambda^{2}UV)\over
\lambda^{2}(1-U)(1-V)}, \nonumber \\[0.2cm]
\omega_{4}&={\sqrt{s}\over 2}{(1+\lambda^{2}U^{2})(1+\lambda^{2}V)\over
\lambda^{2}(U-1)(U-V)}, 
\label{eq:omegas_UV}\\[0.2cm]
\omega_{5}&={\sqrt{s}\over 2}{(1+\lambda^{2}V^{2})(1+\lambda^{2}U)\over
\lambda^{2}(V-1)(V-U)}.\nonumber
\end{align}
We have seen already that in order to keep the amplitude finite we have to exclude $U=0$ and $V=0$ from the physical region. Now,
energy finiteness further demands that $U\neq 1$, $V\neq 1$, and $U\neq V$. By requiring $\omega_{a}\geq 0$ we find that, for example, 
the region $U>0$, $V>0$ has to be considered unphysical as well. Indeed, if this is the case all three numerators in 
\eqref{eq:omegas_UV} are positive whereas the three denominators cannot have the same sign simultaneously. As a consequence, at
least one of the energies has to be negative and the configuration is excluded. Studying the values of $U$ and $V$ in which the three energies
are simultaneously positive for a given $\lambda$, we arrive at the physical regions shown in Fig. \ref{fig:physical_regions}.
Notice that the plot is symmetric under the exchange $U\leftrightarrow V$.

The conformal structure of the equation defining the planar zeros indicates that each solution of Eq. \eqref{eq:cubic_eq} can be realized in infinitely many physical setups, depending on the value of the parameter $\lambda$. 
Notice that the boundaries of the allowed regions depend on $\lambda$ as well, so they move as this parameter varies, while the position of
the zeros, being a projective invariant, remain fixed.  

A particularly interesting regime is the soft limit, in which one or various of the emitted gluon energies tend to zero. From Eq. 
\eqref{eq:omegas_UV} we see that the points in the $(U,V)$ plane for which $\omega_{a}$ vanish are given by
\begin{align}
UV&=-{1\over \lambda^{2}} \hspace*{1cm} (\omega_{3}=0), \nonumber \\[0.2cm]
V&=-{1\over \lambda^{2}} \hspace*{1cm} (\omega_{4}=0), \\[0.2cm]
U&=-{1\over \lambda^{2}} \hspace*{1cm} (\omega_{5}=0), \nonumber 
\end{align}
which are indicated by the dashed lines in Fig. \ref{fig:physical_regions}. On general grounds, a planar zero corresponding
to a point of the cubic \eqref{eq:cubic_eq} can be 
physically captured in the soft limit provided there is a value of $\lambda$ for which 
this point collides against any of the ``soft'' lines defining the boundaries of the physical region. 

\begin{figure}[t]
\centerline{\includegraphics[scale=0.55]{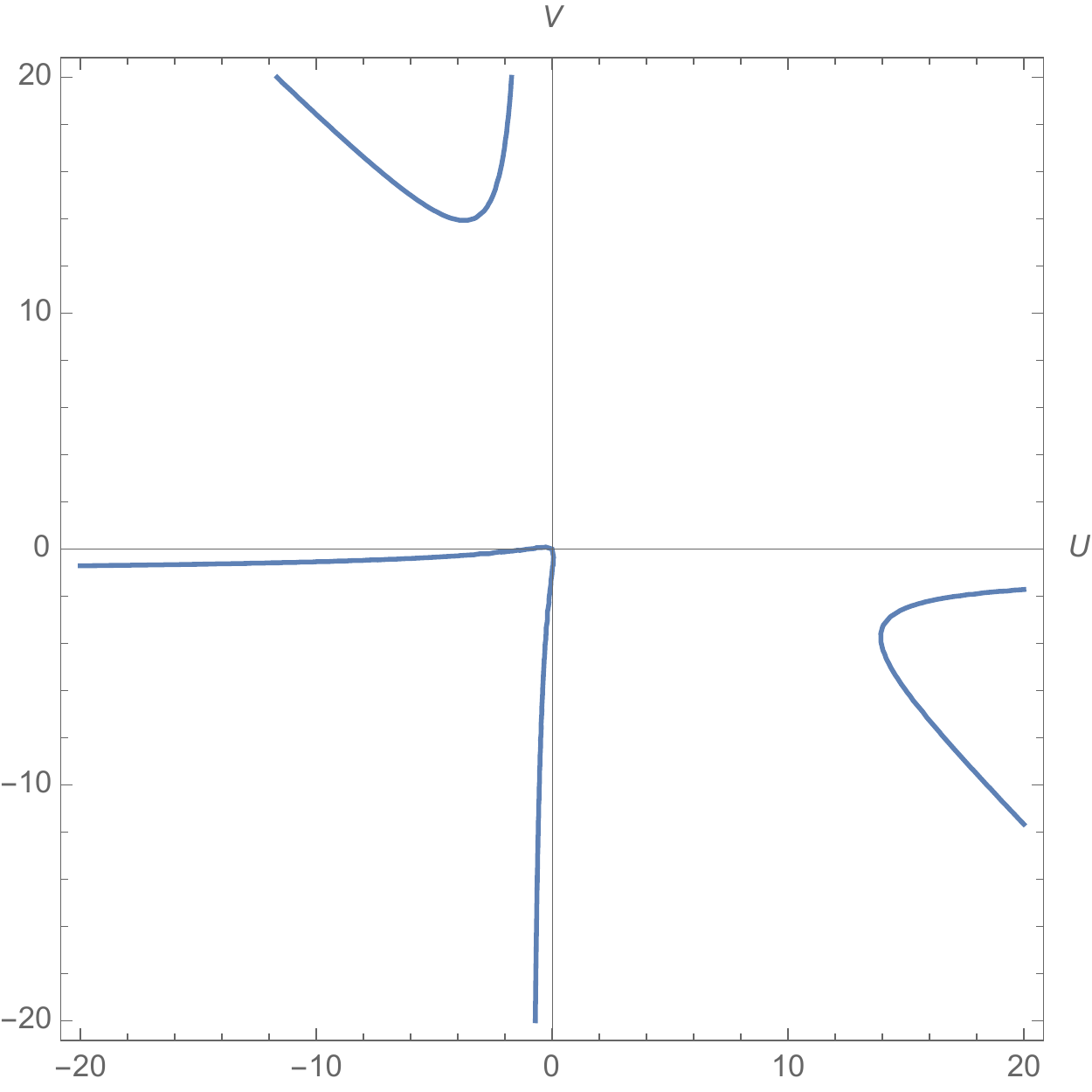}\hspace*{2cm}\includegraphics[scale=0.55]{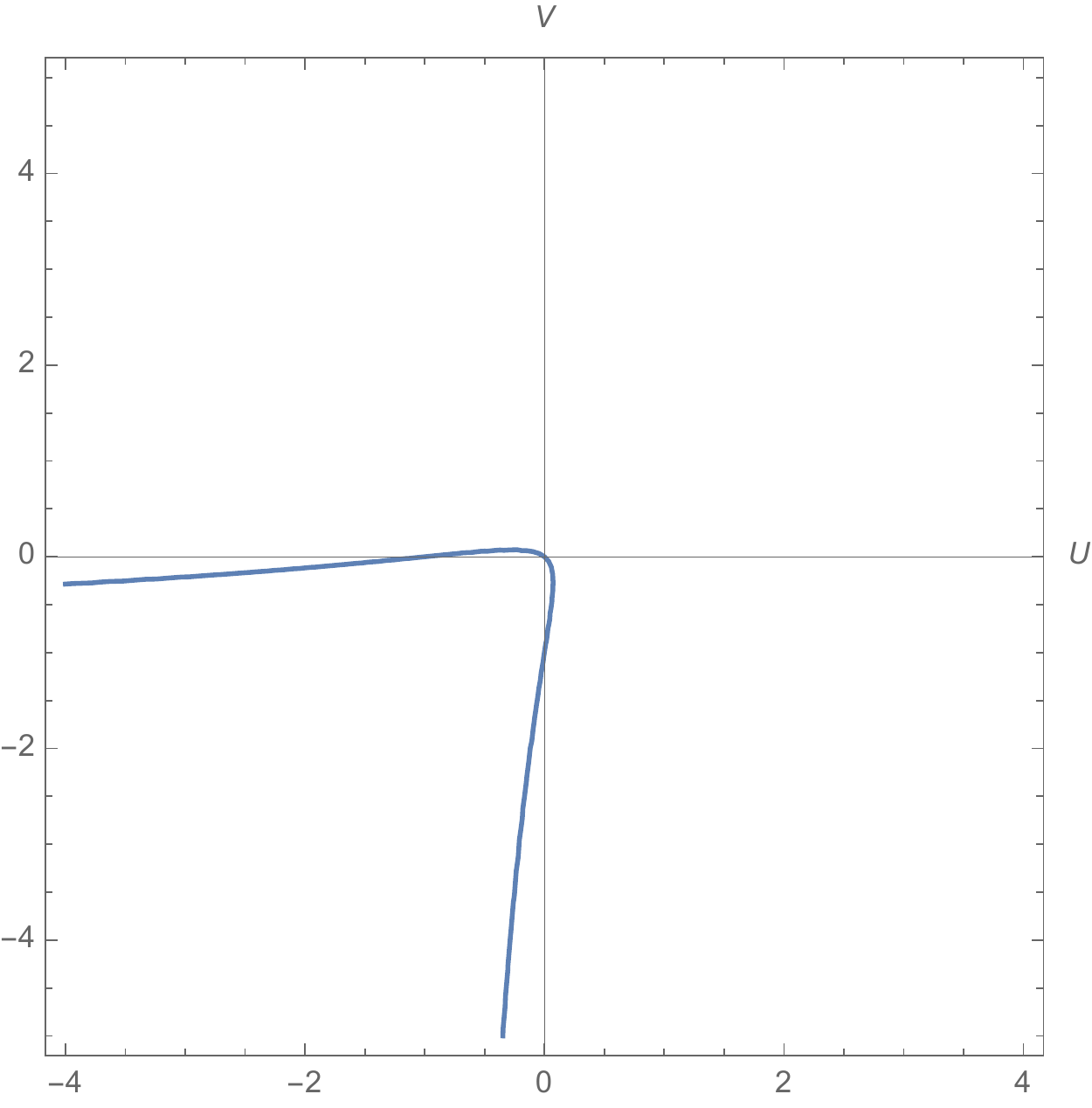}}
\caption[]{Algebraic curve associated with the planar zeros for incoming gluons in a singlet state, Eq. \eqref{eq:cubid_eq_singlet}. 
The right panel shows a blowup of the region 
around $(U,V)=(0,0)$.}
\label{fig:singlet_case}
\end{figure}

The first example to analyze is the case of incoming gluons in a singlet state for arbitrary gauge group, already studied in \cite{harland-lang}. Using
the fact that $f^{da_{3}b}f^{ba_{4}c}f^{ca_{5}d}\sim f^{a_{3}a_{4}a_{5}}$, we find
\begin{align}
c_{2}=c_{6}=-c_{7}=c_{8}=-c_{11}=-c_{13}=-f^{a_{3}a_{4}a_{5}}.
\end{align}
The cubic equation then reads
\begin{eqnarray}
U+V+U^{2}+V^{2}-6 UV +U^{2}V+UV^{2}=0.
\label{eq:cubid_eq_singlet}
\end{eqnarray}
The associated algebraic curve is represented in Fig. \ref{fig:singlet_case}. Comparing with Fig. \ref{fig:physical_regions}
we see that for small enough $\lambda$ there is indeed a large part of the curve lying on physically allowed regions.
In particular, for $\lambda<1$ there are solutions with arbitrarily large $|U|$ and $|V|$.

We study next the loci defined by Eq. \eqref{eq:cubic_eq} for SU($N$) gauge groups with different ranks and various color
configurations:

\paragraph{SU($\mathbf{2}$).} In the case of SU(2) it is easy to write a closed expression for the color factors
\begin{align}
c_{2}&=\delta^{a_{3}a_{4}}\epsilon^{a_{2}a_{5}a_{1}}-\delta^{a_{2}a_{4}}\epsilon^{a_{3}a_{5}a_{1}}, \nonumber \\[0.2cm]
c_{6}&=\delta^{a_{5}a_{3}}\epsilon^{a_{2}a_{4}a_{1}}-\delta^{a_{2}a_{3}}\epsilon^{a_{5}a_{4}a_{1}}, \nonumber \\[0.2cm]
c_{7}&=\delta^{a_{1}a_{4}}\epsilon^{a_{3}a_{5}a_{2}}-\delta^{a_{3}a_{4}}\epsilon^{a_{1}a_{5}a_{2}}, \nonumber \\[0.2cm]
c_{8}&=\delta^{a_{2}a_{5}}\epsilon^{a_{4}a_{1}a_{3}}-\delta^{a_{4}a_{5}}\epsilon^{a_{2}a_{1}a_{3}}, \\[0.2cm]
c_{11}&=\delta^{a_{4}a_{3}}\epsilon^{a_{2}a_{5}a_{1}}-\delta^{a_{2}a_{3}}\epsilon^{a_{4}a_{5}a_{1}}, \nonumber \\[0.2cm]
c_{13}&=\delta^{a_{4}a_{5}}\epsilon^{a_{1}a_{2}a_{3}}-\delta^{a_{1}a_{5}}\epsilon^{a_{4}a_{2}a_{3}}, \nonumber
\end{align}
where a convenient normalization of the gauge group generators has been chosen. 
In principle, the color factors can only take the values $0,\pm 1$, and $\pm 2$,
since each term on the right-hand side of these equations is either $0$ or $\pm 1$. However, the case $\pm 2$ is excluded.
The reason is that due to the structure of indices of the Levi-Civita tensor, sharing the last two entries, 
they cannot have oposite signs. As a consequence, they cannot add up and we conclude that for SU(2) the
color factors satisfy $c_{i}=0,\pm 1$.

An exploration of the possible external color numbers show that there are no solutions containing physical points. We 
illustrate this with a few examples. Our first case has color structure $(a_{1},a_{2},a_{3},a_{4},a_{5})=(2,3,1,1,1)$, 
giving the same value for all color factors
\begin{align}
c_{2}=c_{6}=c_{7}=c_{8}=c_{11}=c_{13}=1.
\end{align}
The resulting cubic equation completely factorizes as
\begin{align}
(U-1)(V-1)(U-V)=0.
\label{eq:curve1_SU(2)}
\end{align}
We see that the three solutions lie outside the physical region and as a consequence there are no planar zeros for this 
gauge configuration.

Next we try $(a_{1},a_{2},a_{3},a_{4},a_{5})=(2,2,2,1,3)$, which corresponds to color factors
\begin{align}
c_{2}=c_{7}=c_{8}=c_{13}=0, \hspace*{1cm} c_{6}=-c_{11}=1.
\end{align}
In this case the equation for the zeros becomes quadratic and factorizes as
\begin{align}
(U-V)^{2}=0.
\end{align}
The geometric loci of zeros coincide again with the unphysical region corresponding to two particles in the final state with
infinite energy. 

As a last example, we take $(a_{1},a_{2},a_{3},a_{4},a_{5})=(1,2,2,2,3)$, which gives
\begin{align}
c_{2}=c_{8}=c_{11}=c_{13}=0, \hspace*{1cm} c_{6}=c_{7}=1.
\end{align}
In this case the cubic again degenerates into a quadratic equation
\begin{align}
U(U-1)=0,
\end{align}
which has no physical solutions.

To summarize, a scan of possible values of the external color numbers show that the only curves obtained in this
case coincide with unphysical regions in the plot of Fig. \ref{fig:physical_regions}, $U=0,1$, $V=0,1$ or $U=V$. 
The only possibility for planar zeros in this case is to consider external states without well-defined color numbers, such
as the singlet case studied above.

\begin{figure}[t]
\centerline{\includegraphics[scale=0.55]{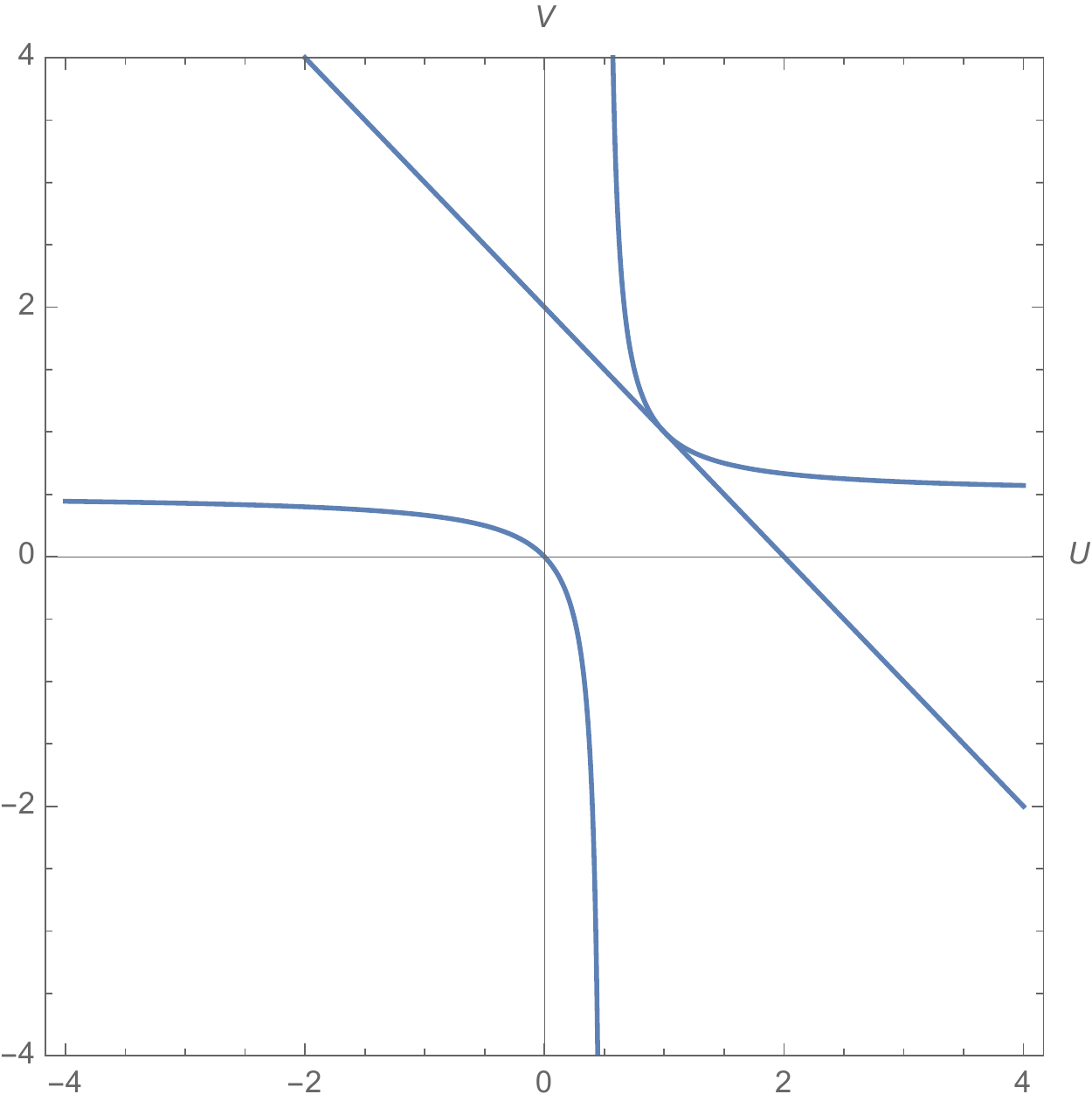}
\hspace*{2cm}\includegraphics[scale=0.55]{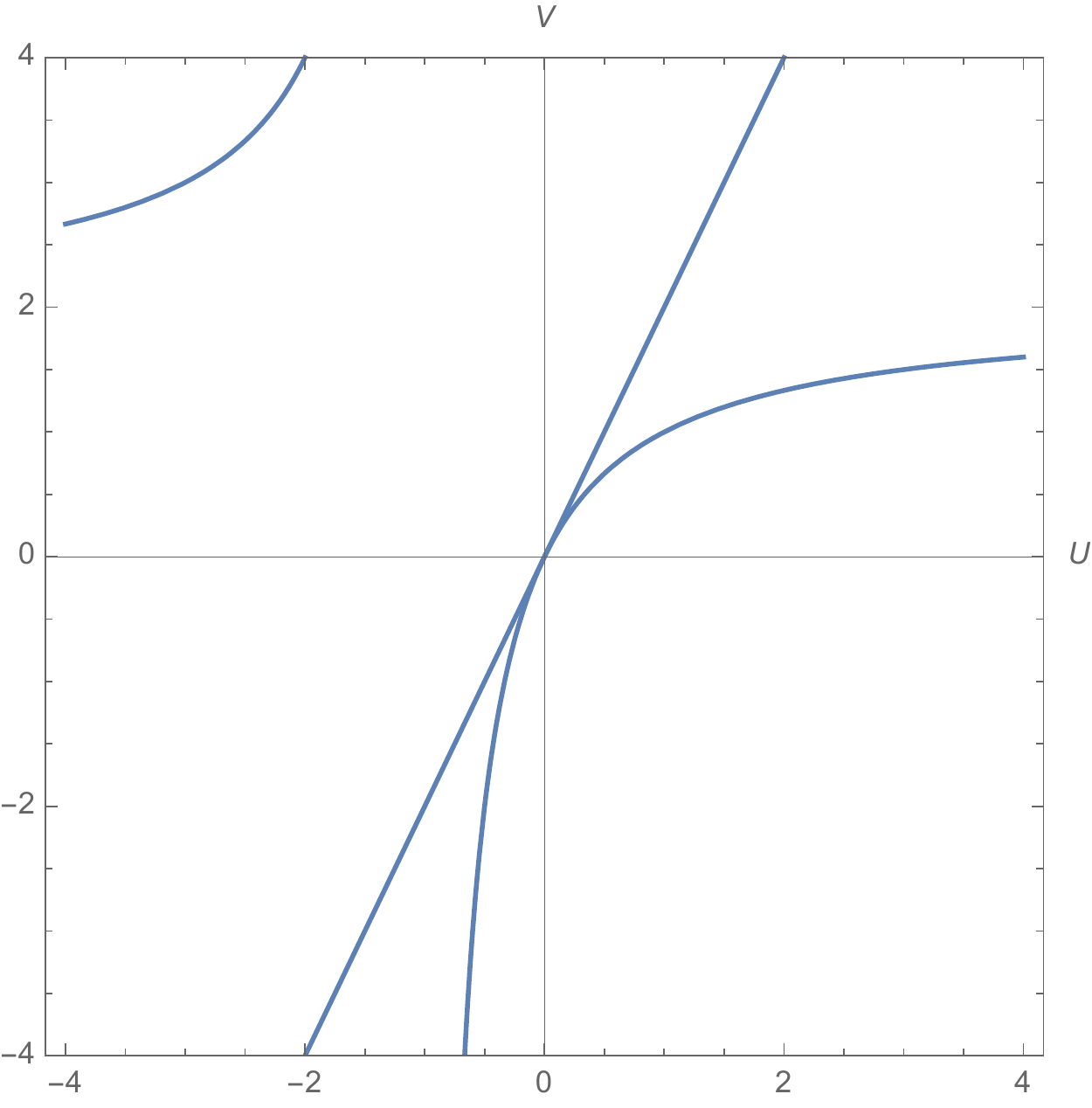}
}
\caption[]{Left panel: curve \eqref{eq:curve_SU(3)_ex1} giving the planar zeros for the SU(3) five-gluon amplitude with 
$(a_{1},a_{2},a_{3},a_{4},a_{5})=(7,7,6,1,5)$. Right panel: the same for $(a_{1},a_{2},a_{3},a_{4},a_{5})=(1,4,1,2,6)$, 
corresponding to Eq. \eqref{eq:second_hyp_SU(2)}.}
\label{fig:SU(3)_case1}
\end{figure}

\paragraph{SU($\mathbf{3}$).} We work out a first example where we take color quantum numbers $(a_{1},a_{2},a_{3},a_{4},a_{5})=(7,7,6,1,5)$ and
color factors
\begin{align}
c_{2}=-c_{7}=c_{8}=-c_{13}=2, \hspace*{1cm} c_{6}=-c_{11}=-1.
\end{align}
The planar zeros are given by the factorized cubic
\begin{align}
(U+V-2)(U+V-2UV)=0.
\label{eq:curve_SU(3)_ex1}
\end{align}
This is a hyperbola together with its tangent at $(U,V)=(1,1)$ (see the left panel of Fig. \ref{fig:SU(3)_case1}). The loci has nonvanishing intersection with
the physically allowed region in the $UV$ plane for appropriate values of $\lambda$. 

A different hyperbola is obtained for $(a_{1},a_{2},a_{3},a_{4},a_{5})=(1,4,1,2,6)$ with
\begin{align}
c_{2}=-c_{11}=-1,\hspace*{1cm} c_{6}=-4, \hspace*{1cm} c_{7}=c_{8}=0, \hspace*{1cm}  c_{13}=-2.
\end{align}
The equation determining the zeros also factorizes in this case, giving
\begin{align}
(2U-V)(-2U+V+UV)=0.
\label{eq:second_hyp_SU(2)}
\end{align}
Again, we have a hyperbola and one of its tangents, this time at the origin. 
The curves are shown in the RHS panel of Fig. \ref{fig:SU(3)_case1}.

As in the SU(2) cases all examples explored for the gauge group SU(3) show factorization of the cubic equation. In this latter case, 
however, not only the type of curves is enlarged to include hyperbolas which were absent for SU(2), but the curves contain
physical points. In addition, considering quantum numbers in a SU(2) 
subgroup of SU(3) generates the curves obtained for the former group. 

\begin{figure}[t]
\centerline{\includegraphics[scale=0.55]{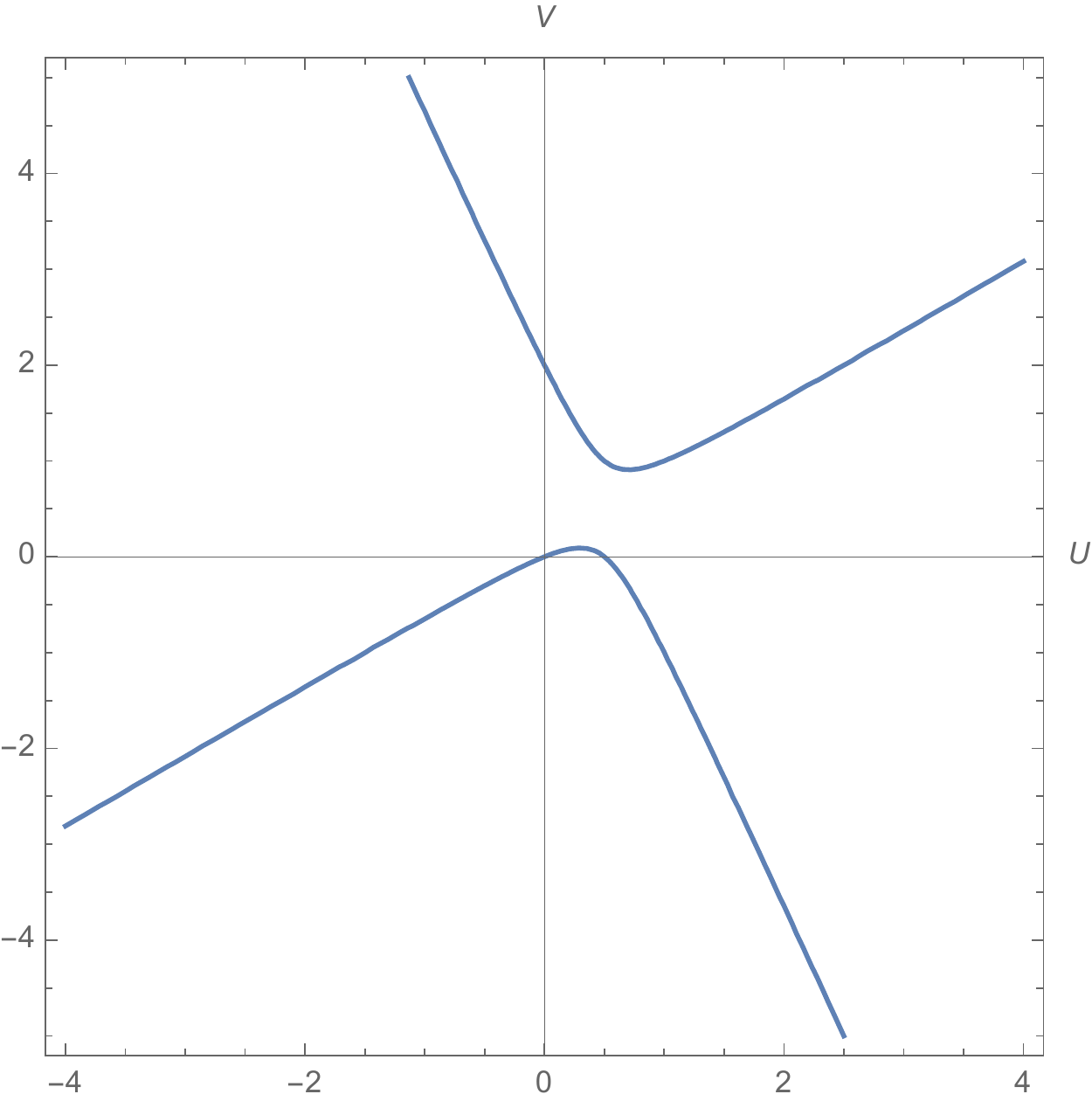}\hspace*{2cm}\includegraphics[scale=0.55]{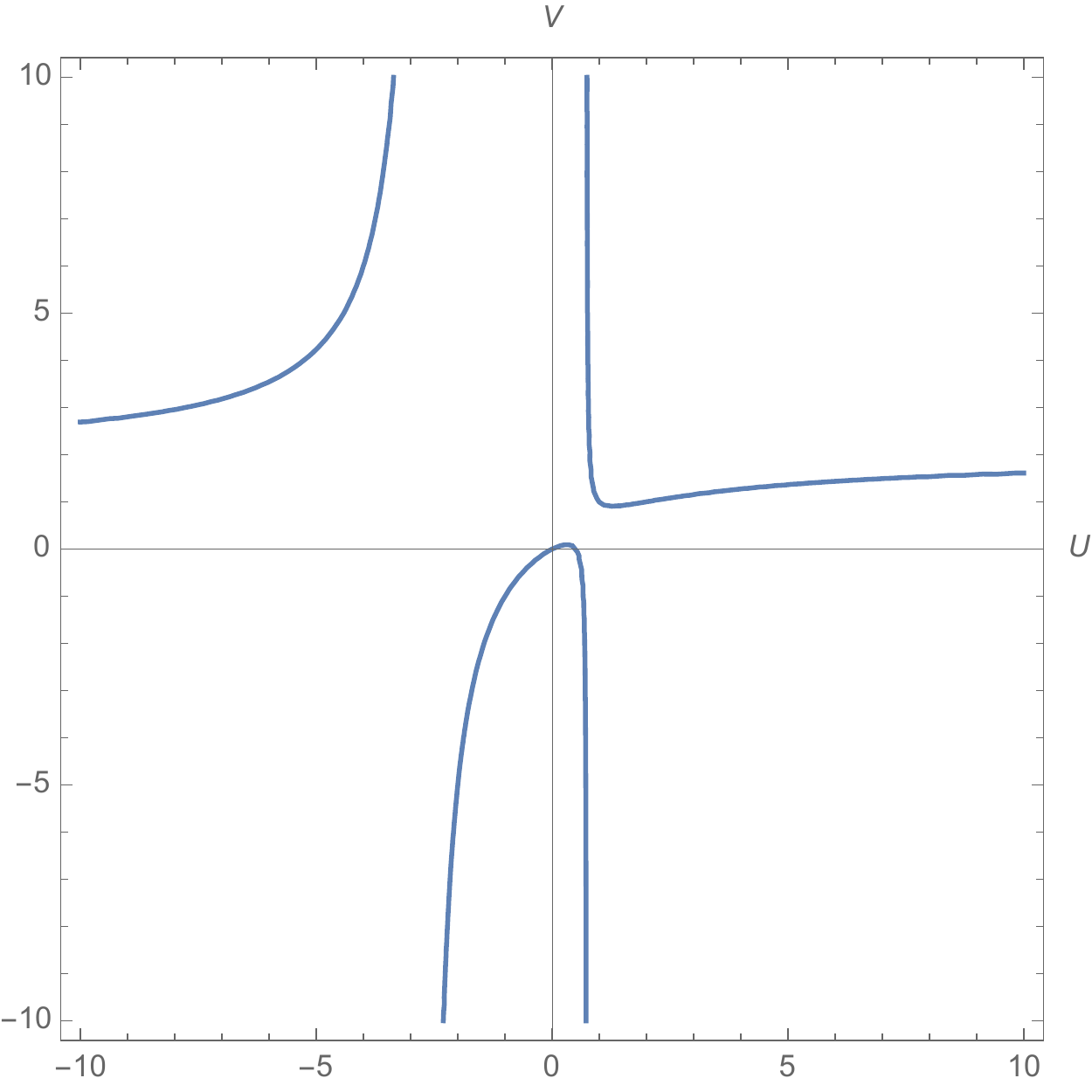}}
\caption[]{Two examples of planar zero curves for the group SU(5): the hyperbola in the LHS panel corresponds to 
$(a_{1},a_{2},a_{3},a_{4},a_{5})=(17,19,19,18,23)$, resulting in Eq. \eqref{eq:alg_curve1_SU(5)}. Equation \eqref{eq:SU(5)curve_ex2} 
is represented on the RHS panel, corresponding to color indices $(a_{1},a_{2},a_{3},a_{4},a_{5})=(19,18,23,17,19)$.}
\label{fig:plotsSU(5)1}
\end{figure}

\paragraph{SU(5).} Enlarging the gauge group to SU(5) brings more general types of cubic algebraic curves. This is for example the case 
taking $(a_{1},a_{2},a_{3},a_{4},a_{5})=(17,19,19,18,23)$. The resulting color factors are
\begin{eqnarray}
c_{2}=c_{13}=0, \hspace*{1cm} c_{6}=c_{8}=2, \hspace*{1cm} c_{7}=c_{11}=1.
\end{eqnarray}
Since $c_{2}$ and $c_{13}$ vanish, it results in the following quadratic equation determining the planar zeros
\begin{eqnarray}
U - 2 U^{2} - 2 V + 2 U V + V^{2}=0.
\label{eq:alg_curve1_SU(5)}
\end{eqnarray}
Unlike the examples encountered for SU(2) and SU(3), this curve does not factorize and corresponds to the hyperbola shown in the LHS panel of Fig. \ref{fig:plotsSU(5)1}.  

A second interesting example is provided by $(a_{1},a_{2},a_{3},a_{4},a_{5})=(19,18,23,17,19)$.
The corresponding color factors are 
\begin{eqnarray}
c_{2}=c_{11}=0, \hspace*{1cm} c_{6}=c_{8}=2, \hspace*{1cm} c_{7}=c_{13}=1.
\end{eqnarray}
The resulting equation for the zero
\begin{eqnarray}
U - 2 U^{2} - 2 V + 2 U V + U^{2} V=0
\label{eq:SU(5)curve_ex2}
\end{eqnarray}
is the cubic curve shown in the RHS panel of Fig. \ref{fig:plotsSU(5)1}. 

\begin{figure}[t]
\centerline{\includegraphics[scale=0.55]{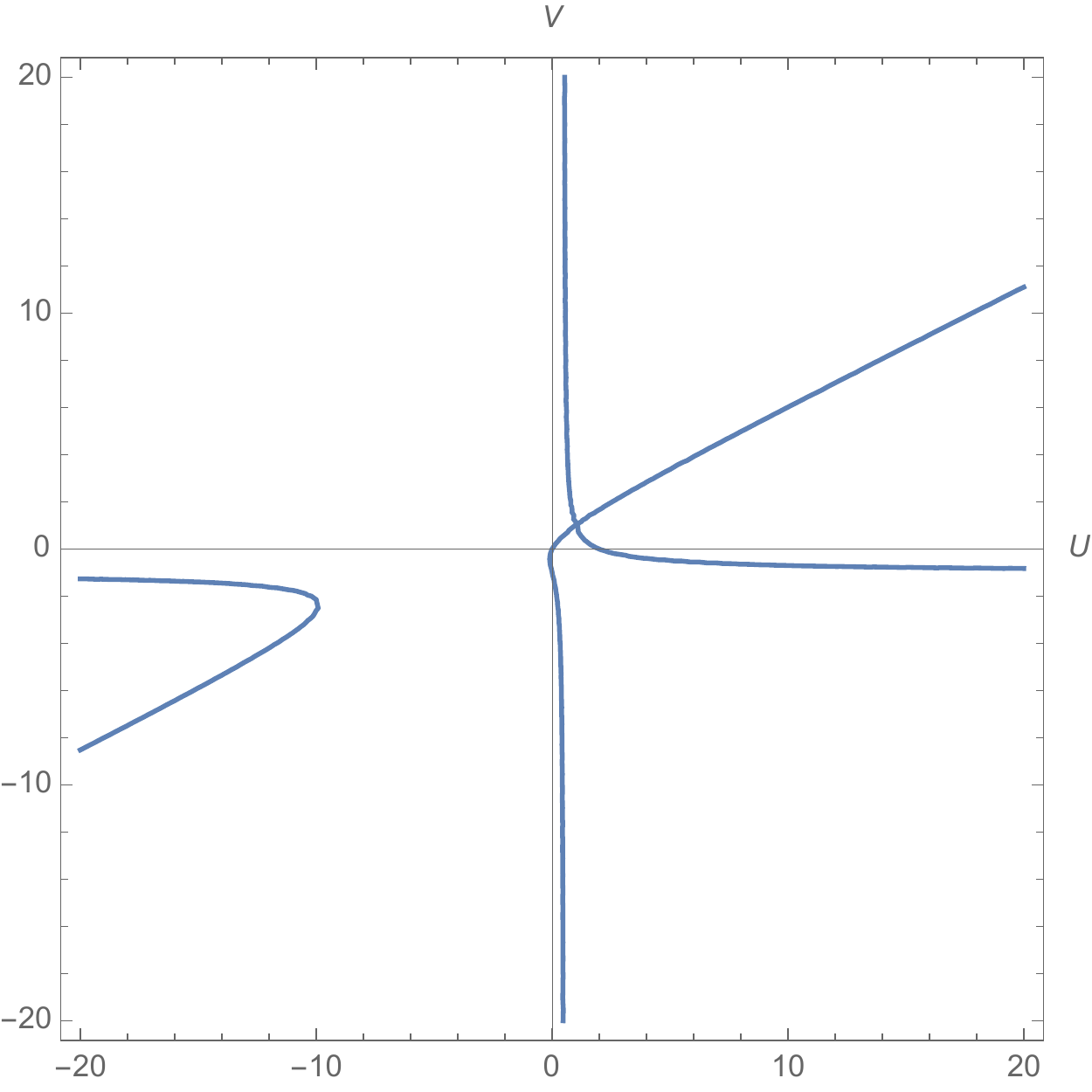}\hspace*{2cm} \includegraphics[scale=0.55]{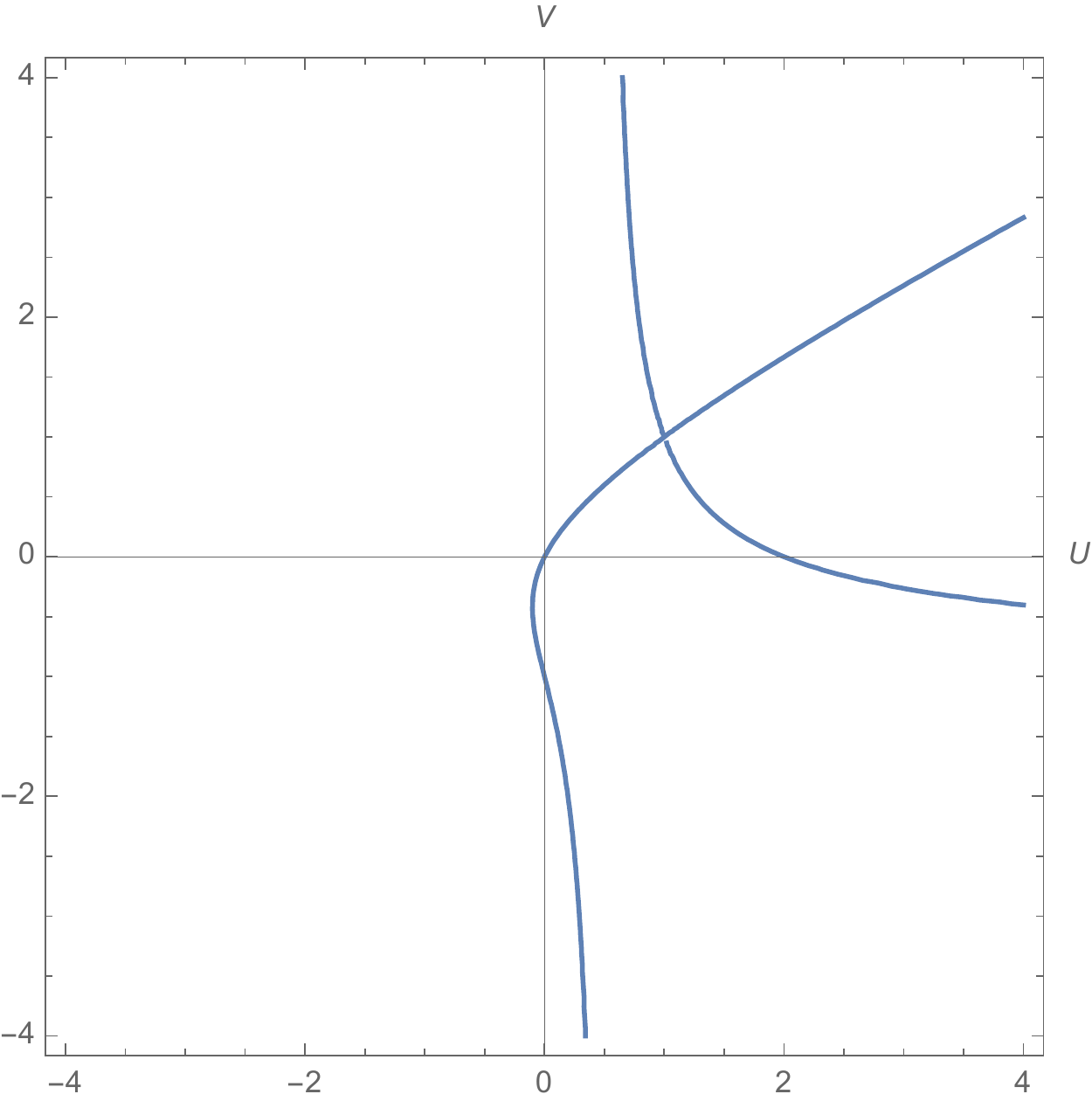}}
\caption[]{Singular curve in Eq. \eqref{eq:curveSU(3)Ex3} 
corresponding to $(a_{1},a_{2},a_{3},a_{4},a_{5})=(19,19,18,23,17)$. The RHS panel blows up the region around 
the origin.}
\label{fig:plotSU(5)Ex3}
\end{figure}

As a last example we take $(a_{1},a_{2},a_{3},a_{4},a_{5})=(19,19,18,23,17)$, with
color factors
\begin{eqnarray}
c_{2}=-c_{7}=-2, \hspace*{1cm} c_{6}=c_{8}=-c_{11}=-c_{13}=1.
\end{eqnarray}
We get the cubic curve
\begin{eqnarray}
2U-U^{2}-V-U^{2}V-V^{2}+2UV^{2}=0,
\label{eq:curveSU(3)Ex3}
\end{eqnarray}
which, as shown in Fig. \ref{fig:plotSU(5)Ex3}, contains a singular point at $(U,V)=(1,1)$.

We see how SU(5) provides more general types of curves than the ones found for unitary groups of lower rank. We also have
to take into account that SU(5) contains SU(3) and SU(2) subgroups. Using the standard generators (see, for example, \cite{mohapatra})
these subgroups are respectively generated by $\{T^{1},\ldots,T^{8}\}$ and $\{T^{21},
T^{22},T^{23}\}$. Thus, setting the external indices in the subsets $(1,\ldots,8)$ or $(21,22,23)$ we recover previous examples. 
For instance, $(a_{1},a_{2},a_{3},a_{4},a_{5})=(7,7,6,1,5)$ gives the curve shown in the LHS panel of Fig. \ref{fig:SU(3)_case1}, 
whereas $(a_{1},a_{2},a_{3},a_{4},a_{5})=(22,23,21,21,21)$ reproduces Eq. \eqref{eq:curve1_SU(2)}.

\section{Planar zeros and color permutations}
\label{sec:permutations}

It is interesting to see how the zeros here investigated transform under permutations of the color quantum numbers of the external particles. We begin considering 
those permutations preserving the choice of amplitudes basis implied by Eq. \eqref{eq:kleiss-kuijf_exp}. These are elements of $S_{3}$ permuting
the color indices of the three outgoing gluons (see Fig. \ref{fig:KK-rel}). 

In order to find the action of these permutations on the geometric loci of planar zeros, we can see how the color factors 
\eqref{eq:color_factors} transform under permutations of the $(a_{3},a_{4},a_{5})$ color indices. Here instead we use a more
geometric approach and work with the homogenization of the cubic equation \eqref{eq:cubic_eq}
\begin{align}
c_{7}Z^{2}U-c_{8}Z^{2}V-c_{6}ZU^{2}+c_{11}ZV^{2}+(c_{2}+c_{6}-c_{7}+c_{8}-c_{11}-c_{13})ZUV
\nonumber \\[0.2cm]
+c_{13}U^{2}V-c_{2}UV^{2}=0.
\label{eq:cubic_eq_hom}
\end{align}
The group $S_{3}$ acts passively by permutation of the coordinates $(Z,U,V)$. Let us discuss the geometrical meaning of these transformations.
Equation \eqref{eq:cubic_eq_hom} is defined in the whole projective plane, which is covered by the three affine patches centered at $(1,0,0)$, 
$(0,1,0)$, and $(0,0,1)$. The group $S_{3}$ is generated by $S_{2}$ transformations interchanging the two coordinates within each patch, together
with cyclic permutations of the three patches. This defines a coset decomposition of $S_{3}$ with respect to its cyclic subgroup generated by
$(123)$.

The physically allowed regions shown in Fig. \ref{fig:physical_regions} correspond to the patch centered at 
$(1,0,0)$. It has been already pointed out that it is invariant under the interchange of the two coordinates $U\leftrightarrow V$.
Moreover, the corresponding plots in the other two affine coordinate patches are identical to this one. This is easy to see from Eq. \eqref{eq:energies_dehomog}, where it is glaring that cyclic permutations of the three patches only interchange the energies of the three 
outgoing gluons. Thus, 
the positivity conditions remain algebraically the same in any of the three affine patches. The final conclusion is that 
$S_{3}$ only acts on the axes labels of the plot in Fig. \ref{fig:physical_regions}. This is a passive version of the fact that the energies
of the outgoing gluons are determined by momentum conservation alone and that the color structures play no role in it.
  
Applying a permutations of $S_{3}$ to Eq. \eqref{eq:cubic_eq_hom}, we find that the color factors 
transform under the (six-dimensional) regular
representation of the group: in particular, if we write $\boldsymbol{C}=(c_{2},c_{6},c_{7},c_{8},c_{11},c_{13})^{T}$ 
the group $S_{3}$ acts through the matrices
\begin{align}
(1)(2)(3)&=\left(
\begin{array}{cccccc}
1 & 0 & 0 & 0 & 0 & 0 \\
0 & 1 & 0 & 0 & 0 & 0 \\
0 & 0 & 1 & 0 & 0 & 0 \\
0 & 0 & 0 & 1 & 0 & 0 \\
0 & 0 & 0 & 0 & 1 & 0 \\
0 & 0 & 0 & 0 & 0 & 1
\end{array}
\right), \hspace*{1.3cm}
(123)= \left(
\begin{array}{cccccc}
0 & 0 & 0 & 1 & 0 & 0 \\
1 & 0 & 0 & 0 & 0 & 0 \\
0 & 0 & 0 & 0 & 0 & 1 \\
0 & 1 & 0 & 0 & 0 & 0 \\
0 & 0 & 1 & 0 & 0 & 0 \\
0 & 0 & 0 & 0 & 1 & 0
\end{array}
\right), \nonumber \\[0.2cm]
(132)&=\left(
\begin{array}{cccccc}
0 & 1 & 0 & 0 & 0 & 0 \\
0 & 0 & 0 & 1 & 0 & 0 \\
0 & 0 & 0 & 0 & 1 & 0 \\
1 & 0 & 0 & 0 & 0 & 0 \\
0 & 0 & 0 & 0 & 0 & 1 \\
0 & 0 & 1 & 0 & 0 & 0
\end{array}
\right),\hspace*{1cm}
(12)(3)= \left(
\begin{array}{cccccc}
0 & 0 & 0 & 0 & 1 & 0 \\
0 & 0 & 1 & 0 & 0 & 0 \\
0 & 1 & 0 & 0 & 0 & 0 \\
0 & 0 & 0 & 0 & 0 & 1 \\
1 & 0 & 0 & 0 & 0 & 0 \\
0 & 0 & 0 & 1 & 0 & 0
\end{array}
\right), \hspace*{1cm} \label{eq:regular_rep_S3}\\[0.2cm]
(13)(2)&=\left(
\begin{array}{cccccc}
0 & 0 & 1 & 0 & 0 & 0 \\
0 & 0 & 0 & 0 & 0 & 1 \\
1 & 0 & 0 & 0 & 0 & 0 \\
0 & 0 & 0 & 0 & 1 & 0 \\
0 & 0 & 0 & 1 & 0 & 0 \\
0 & 1 & 0 & 0 & 0 & 0
\end{array}
\right),\hspace*{1cm}
(1)(23)= \left(
\begin{array}{cccccc}
0 & 0 & 0 & 0 & 0 & 1 \\
0 & 0 & 0 & 0 & 1 & 0 \\
0 & 0 & 0 & 1 & 0 & 0 \\
0 & 0 & 1 & 0 & 0 & 0 \\
0 & 1 & 0 & 0 & 0 & 0 \\
1 & 0 & 0 & 0 & 0 & 0
\end{array}
\right). \nonumber
\end{align}
Notice that the combination of color factors in the coefficient of the $ZUV$ term itself transforms with the one-dimensional 
parity representation, 
\begin{align}
\sigma(c_{2}+c_{6}-c_{7}+c_{8}-c_{11}-c_{13})=(-1)^{\pi(\sigma)}(c_{2}+c_{6}-c_{7}+c_{8}-c_{11}-c_{13}),
\end{align}
where $\pi(\sigma)=0,1$ for even and odd permutations respectively. This is a consequence of the fact that $ZUV$ is an invariant under
the permutation group.

Applying the transformations given by the matrices \eqref{eq:regular_rep_S3} to the curve \eqref{eq:cubid_eq_singlet}  obtained in the case of  
the scattering of two gluons in a singlet state, $\boldsymbol{C}=(1,1,-1,1,-1,-1)^{T}$, we find that it is invariant, since
$\boldsymbol{C}$ transforms with the parity of the permutation, 
$\sigma(\boldsymbol{C})=(-1)^{\pi(\sigma)}\boldsymbol{C}$. This means the curve shown in the plot 
in Fig. \ref{fig:singlet_case} describes the planar zeros in all three coordinate patches. Incidentally, notice that 
the curve is invariant as well under the interchange of the two coordinates in any of the three corresponding plots. 

The curves presented in Section \ref{sec:gluon_planar} are expressed in the patch $(1,0,0)$. Under permutation of the two coordinates, some
solutions remain invariant, such as Eq. \eqref{eq:curve_SU(3)_ex1}, or get mapped into a different solution.

We can also consider the transformation of the curves with respect to general color permutations. They form the group $\mathsf{TCS}_{5}$
which is identified with the cyclic Lie operad $\mathsf{Lie}(\!(5)\!)$. Its structure has been studied in Ref. \cite{kol_shir}, where it was
found that its action on the six independent color structures is given by the following six-dimensional representation of $S_{5}$
\begin{align*}
\yng(3,1,1)\,\,\,.
\end{align*}
Unlike the transformations of $S_{3}$ studied above, those in $\mathsf{TCS}_{5}\backslash S_{3}$ do not act on the geometric loci of planar zeros
by permutation of the projective coodinates $(Z,U,V)$.
The most obvious example is provided by the interchange of the color indices of the incoming gluons $a_{1}$, $a_{2}$. From Eq. \eqref{eq:color_factors}
we find that this transformation acts linearly on $\boldsymbol{C}$ through the matrix
\begin{align}
(12)(3)(4)(5)=\left(
\begin{array}{rrrrrr}
0 & 0 & -1 & 0 & 0 & 0 \\
0 & 0 & 0 & 0 & -1 & 0 \\
-1 & 0 & 0 & 0 & 0 & 0 \\
0 & 0 & 0 & 0 & 0 & -1 \\
0 & -1 & 0 & 0 & 0 & 0 \\
0 & 0 & 0 & -1 & 0 & 0
\end{array}
\right),
\end{align}
where we have used the cycle notation for the elements of $S_{5}$. It is interesting to point out that this transformation leaves invariant 
the coefficient of the $ZUV$ coefficient in \eqref{eq:cubic_eq_hom}. This last property is not shared by other 
transformations in $\mathsf{TCS}_{5}\backslash S_{3}$. For example, 
\begin{align}
(134)(25)=\left(
\begin{array}{rrrrrr}
-1 & 0 & 0 & 0 & 0 & 1 \\
0 & 0 & 1 & 0 & 0 & 0 \\
0 & -1 & 1 & 0 & 0 & 0 \\
-1 & 0 & 0 & 0 & 1 & 0 \\
-1 & 0 & 0 & 0 & 0 & 0 \\
0 & 0 & 1 & -1 & 0 & 0
\end{array}
\right), \hspace*{0.5cm}
(1245)(3)=\left(
\begin{array}{rrrrrr}
-1 & 1 & -1 & 0 & 0 & 0 \\
0 & 1 & 0 & 0 & 0 & 0 \\
0 & 0 & 0 & 0 & 0 & 1 \\
-1 & 0 & 0 & 0 & 0 & 1 \\
-1 & 0 & -1 & 1 & 0 & 1 \\
0 & 1 & -1 & 0 & 0 & 0
\end{array}
\right),
\end{align}
act on $c_{2}+c_{6}-c_{7}+c_{8}-c_{11}-c_{13}$ respectively as
\begin{align}
(134)(25)(c_{2}+c_{6}-c_{7}+c_{8}-c_{11}-c_{13})&=c_{2}-c_{6}+c_{7}-c_{8}-c_{11}-c_{13}, \nonumber \\[0.2cm]
(1245)(3)(c_{2}+c_{6}-c_{7}+c_{8}-c_{11}-c_{13})&=c_{2}-c_{6}-c_{7}+c_{8}-c_{11}+c_{13}.
\end{align}

Through its transformations of the color factors, $\mathsf{TCS}_{5}$ acts on the curves determining the planar zeros. 
In fact, this action can be used to generate the whole orbit of projective curves associated with the permutations of the
color quantum numbers of the interacting gluons. 

\section{Graviton planar zeros from color-kinematics duality}
\label{sec:graviton}

We turn now to the problem of planar zeros in the five-graviton tree level amplitude. The gravitational amplitude can be constructed 
from its gluon counterpart \eqref{eq:amplitude_gluon_general} using the BCJ prescription \cite{color-kinematics}, 
\begin{align}
-i\mathcal{M}_{5}&=\left({\kappa\over 2}\right)^{3}\left({n_{1}^{2}\over s_{12}s_{45}}+{n_{2}^{2}\over s_{23}s_{15}}+{n_{3}^{2}\over s_{34}s_{12}}
+{n^{2}_{4}\over s_{45}s_{23}}+{n_{5}^{2}\over s_{15}s_{34}}+{n_{6}^{2}\over s_{14}s_{25}}
+{n_{7}^{2}\over s_{13}s_{25}}+{n_{8}^{2}\over s_{24}s_{13}} \right. \nonumber \\[0.2cm]
&+ \left.{n_{9}^{2}\over s_{35}s_{24}}+{n_{10}^{2}\over s_{14}s_{35}}+{n_{11}^{2}\over s_{15}s_{24}}
+{n_{12}^{2}\over s_{12}s_{35}}+{n_{13}^{2}\over s_{23}s_{14}}+{n_{14}^{2}\over s_{25}s_{34}}
+{n_{15}^{2}\over s_{13}s_{45}}\right),
\end{align}
provided the numerators $n_{i}$ satisfy color-kinematics duality, with $\kappa$ the gravitational coupling. 
Taking the graviton polarizations $(1^{-},2^{-},3^{+},4^{+},5^{+})$, we use our expression 
for the MHV gauge amplitude given in Eq. \eqref{eq:5g_amplitude_spinors} to write
\begin{align}
-i\mathcal{M}_{5}
&=-i\left({\kappa\over 2}\right)^{3}\langle 12\rangle^{3}\left({n_{2}\over \langle 23\rangle\langle 34\rangle\langle 45\rangle\langle 51\rangle}
+{n_{6}\over \langle 25\rangle\langle 53\rangle\langle 34\rangle\langle 41\rangle}
+{n_{7}\over \langle 25\rangle\langle 53\rangle\langle 43\rangle\langle 31\rangle}\right. \nonumber \\[0.2cm]
&\left.+{n_{8}\over \langle 24\rangle\langle 45\rangle\langle 53\rangle\langle 31\rangle} 
+{n_{11}\over \langle 24\rangle\langle 43\rangle\langle 35\rangle\langle 51\rangle} 
+{n_{13}\over \langle 23\rangle\langle 35\rangle\langle 54\rangle\langle 41\rangle}\right).
\end{align}
Using the form of the gauge theory numerators in Eq. \eqref{eq:numerators}, after a bit of 
algebra we arrive at the simpler expression \cite{bgk}
\begin{align}
-i\mathcal{M}_{5}&=-\left({\kappa\over 2}\right)^{3}{\langle 12\rangle^{7}[41][52]\over \langle 12\rangle\langle 14\rangle
\langle 23\rangle \langle 25\rangle \langle 34\rangle \langle 35\rangle \langle 45\rangle}
\left(1-{\langle 14\rangle\langle 25\rangle [42][51]\over \langle 15\rangle \langle 24\rangle[41][52]}\right).
\end{align}

The term inside the parenthesis can be further simplified taking into account the relation $s_{ij}=\langle ij\rangle[ji]$,
\begin{align}
1-{\langle 14\rangle\langle 25\rangle [42][51]\over \langle 15\rangle \langle 24\rangle[41][52]}
=1-\left({\langle 14\rangle\langle 25\rangle\over \langle 15\rangle\langle 24\rangle}\right)^{2}
\left({s_{15}s_{24}\over s_{14}s_{25}}\right),
\end{align}
which in turn can be written as a function of $\Delta \phi_{45}$, the difference of azimuthal angles of particles
$4$ and $5$ (see the Appendix of Ref. \cite{harland-lang}),
\begin{align}
\left({\langle 14\rangle\langle 25\rangle\over \langle 15\rangle\langle 24\rangle}\right)^{2}
\left({s_{15}s_{24}\over s_{14}s_{25}}\right)=e^{2i\Delta\phi_{45}}.
\end{align}
With this, the five-graviton tree level amplitude reads
\begin{align}
-i\mathcal{M}_{5}&=-\left({\kappa\over 2}\right)^{3}{\langle 12\rangle^{7}[41][52]\over \langle 12\rangle\langle 14\rangle
\langle 23\rangle \langle 25\rangle \langle 34\rangle \langle 35\rangle \langle 45\rangle}
\left(1-e^{2i\Delta\phi_{45}}\right).
\label{eq:grav_ampl}
\end{align}

Given our choice of reference frame, planarity implies that for any two outgoing momenta their azimuthal angles must satisfy
$\Delta\phi_{ij}=0,\pi$. In both cases we find from \eqref{eq:grav_ampl} that the gravitational amplitude vanishes
\begin{align}
-i\mathcal{M}_{5}\Big|_{\rm planar}=0. 
\end{align}
Unlike the gauge case where the cancellation condition depends on the color factors of the incoming particles, 
the graviton amplitude automatically vanishes in the limit of planar scattering. 

We now show that this is a consequence of color kinematic duality. In gauge theories we have seen that the vanishing of the 
amplitude in the planar case leads to a nontrivial condition on the momenta of the outgoing particles given in Eq. 
\eqref{eq:planar_cond1}. In fact, the numerators in Eq. \eqref{eq:numerators} have been chosen to satisfy color-kinematics duality, so
we can obtain the planar zero condition for gravity by replacing the color factors $\{c_{2},c_{6},c_{7},c_{8},c_{11},c_{13}\}$ 
in \eqref{eq:planar_cond1} with the 
corresponding numerators $\{n_{2},n_{6},n_{7},n_{8},n_{11},n_{13}\}$. In terms of the stereographic coordinates, the latter are given by
\begin{align}
n_{6}&=n_{7}=is^{3\over 2}{(\zeta_{3}-\zeta_{5})\zeta_{5}\over \zeta_{3}(1+\zeta_{4}\zeta_{5})}, \nonumber \\[0.2cm]
n_{8}&=is^{3\over 2}{(\zeta_{3}-\zeta_{5})\zeta_{4}\over \zeta_{3}(1+\zeta_{4}\zeta_{5})}, \\[0.2cm]
n_{2}&=n_{11}=n_{13}=0. \nonumber 
\end{align}
After this substitution, the condition for the existence of a zero in the amplitude is identically satisfied 
\begin{align}
-n_{2}{\zeta_{5}-\zeta_{3}\over \zeta_{3}}
-n_{6}{\zeta_{4}-\zeta_{5}\over \zeta_{5}}+n_{7}{\zeta_{3}-\zeta_{5}\over \zeta_{5}}
-n_{8}{\zeta_{3}-\zeta_{4}\over \zeta_{4}}
+n_{11}{\zeta_{5}-\zeta_{4}\over\zeta_{4}}+n_{13}{\zeta_{4}-\zeta_{3}\over \zeta_{3}}
\nonumber \\[0.2cm]
=is^{3\over 2}{(\zeta_{3}-\zeta_{5})\over \zeta_{3}(1+\zeta_{4}\zeta_{5})}\Big(-\zeta_{4}+\zeta_{5}+\zeta_{3}-\zeta_{5}-
\zeta_{3}+\zeta_{4}\Big)=0.
\label{eq:planar_cond1_grav}
\end{align}
This implies that, in the gravitational case, planarity is enough to make the amplitude vanish, without additional kinematic conditions to be satisfied by the stereographic 
coordinates of the outgoing gravitons.

\section{Closing remarks}
\label{sec:conclusions}

\begin{figure}[t]
\centerline{\includegraphics[scale=1.0]{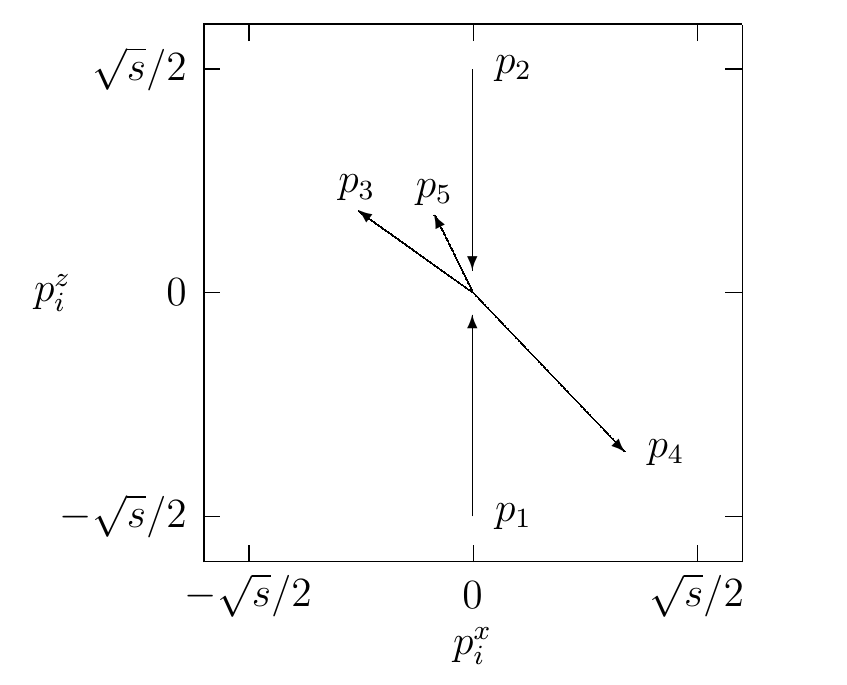}\hspace*{0.5cm}\includegraphics[scale=1.0]{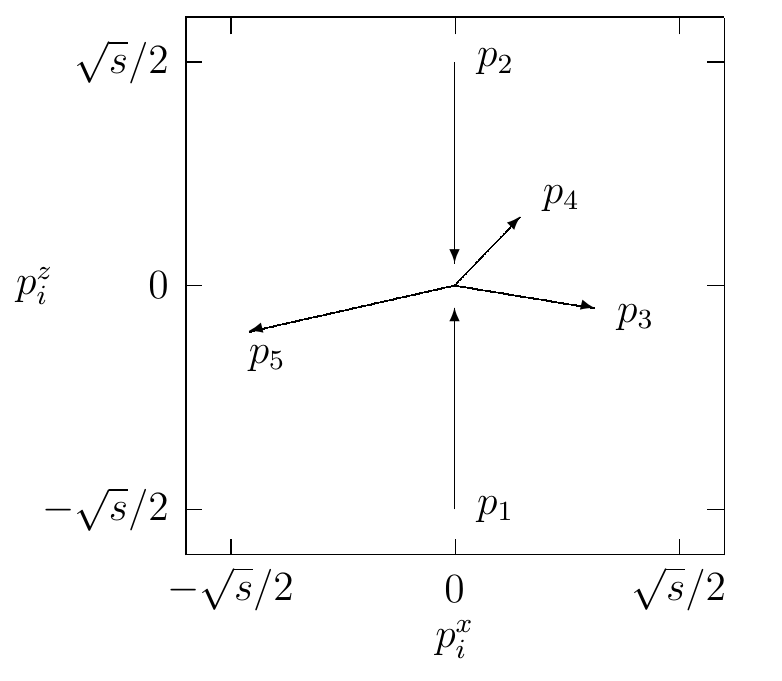}}
\caption[]{Graphic representation of the kinematic configuration on the interaction plane for two planar zeros. 
In both cases the gauge group is SU(3) with color configuration $(7,7,6,1,5)$, so they correspond to two points
on the curve shown on the left panel of \ref{fig:SU(3)_case1}. The left panel represents the solution
$(\zeta_{3},\zeta_{4},\zeta_{5})=(-1.95,0.4,-4.3)$, whereas in the right panel 
$(\zeta_{3},\zeta_{4},\zeta_{5})=(0.85,2.5,-0.8)$. In the projective coordinates \eqref{eq:projective_coord}, 
they correspond to
$\lambda=-1.95$, $U=-0.21$, $V=2.21$ and $\lambda=0.85$, $U=2.94$, $V=-0.94$ respectively.}
\label{fig:vector_plots}
\end{figure}

We have studied the presence of planar zeros in both Yang-Mills theories and gravity. For the case of gauge theories,
we have represented in Fig. \ref{fig:vector_plots} the kinematics on the interaction plane for two typical planar zeros
within the same color configuration. By varying the value of $\lambda\equiv \zeta_{3}$ while keeping $U\equiv\zeta_{4}/\zeta_{3}$
and $V\equiv \zeta_{5}/\zeta_{3}$ constant, these processes can be deformed into 
a different one with the emission, for example, of one or more soft gluons while the total amplitude remains equal to zero.
This happens because planar zeros live in the projective $U$-$V$ plane and are therefore 
invariant under a simultaneous rescaling of the three outgoing stereographic coordinates.

Without loss of generality 
we considered the situation in which the scattering takes place in the $y=0$ plane.
Planar zeros on a different interaction plane can
be obtained by applying rotations to the solutions studied here. In particular, the Lorentz group acts on the 
stereographic coordinates parametrizing the direction of the momenta through $\mbox{SL}(2,\mathbb{C})$ 
transformations \cite{strominger_et_al}
\begin{align}
\zeta_{k}'={a\zeta_{k}+b\over c\zeta_{k}+d}, \hspace*{2cm} ad-bc=1,
\end{align} 
where for the incoming particles we have $\zeta_{1}=\infty$ and $\zeta_{2}=0$. 
Rotations can be spotted by looking for transformations leaving invariant the energies \eqref{eq:energies_dehomog},
together with those of the incoming particles. They are given by 
\begin{align}
\left(
\begin{array}{cc}
a & b \\[0.2cm]
c & d
\end{array}
\right)=\left(
\begin{array}{cc}
\xi & -\sqrt{1-\xi^{2}} \\[0.2cm]
\sqrt{1-\xi^{2}} & \xi
\end{array}
\right).
\end{align}
For real $|\xi|\leq 1$, we parametrize $\xi=\cos\phi$. This corresponds to a rotation of the interaction plane of
angle $2\phi$ with respect to the $x$-axis. Alternatively, for $|\xi|>1$, setting $|\xi|=\cosh{\chi}$ the transformation implements
a rotation of angle $\sin\phi'=\tanh{2\chi}$ around the $y$-axis. 

With the results here presented we have shed some light on the origin of the planar zeroes present in Yang-Mills scattering amplitudes. Our results can be generalized to an arbitrary number of external legs at Born level.  It will be worth further investigating the effect of quantum corrections. We have also connected, via the BCJ duality, these zeroes to the corresponding ones in gravity. We are currently studying how this picture is modified when the scattering of open and closed strings is considered.

\section*{Acknowledgments}

We would like to thank Lucian Harland-Lang and Lev Lipatov for discussions. A.S.V. and D. M. J. acknowledge support from the Spanish Government grant FPA2015-65480-P 
and Spanish MINECO Centro de Excelencia Severo Ochoa Programme (SEV-2012-0249). 
The work of M.A.V.-M. has been partially supported by Spanish Government grant FPA2015-64041-C2-2-P.

\end{document}